\documentstyle[prl,aps,multicol,epsf]{revtex}
\renewcommand{\narrowtext}{\begin{multicols}{2} \global\columnwidth20.5pc}
\renewcommand{\widetext}{\end{multicols} \global\columnwidth42.5pc} 

\begin{document}

\newcommand{\be}{\begin{equation}}
\newcommand{\ee}{\end{equation}}
\newcommand{\bea}{\begin{eqnarray}}
\newcommand{\eea}{\end{eqnarray}}
\newcommand{\nt}{\narrowtext}
\newcommand{\wt}{\widetext}

\title{Quantum impurity models of noisy qubits}

\author{D. V. Khveshchenko}

\address{Department of Physics and Astronomy, University of North
Carolina, Chapel Hill, NC 27599}
\maketitle

\begin{abstract}
We demonstrate that the problem of coupled two-level systems
("qubits") which are also subject to a generic (sub)Ohmic dissipative environment 
belongs to the same class of models as those describing (non)magnetic 
impurities embedded in strongly correlated systems. 
A further insight into the generalized single- and two-impurity Bose/Fermi Kondo
models enables one to make specific recommendations towards 
a systematic engineering of highly coherent multi-qubit assemblies for potential 
applications in quantum information processing.
\end{abstract}

\nt 
The studies of open quantum systems
have long been focused on the behavior of a single two-level system
(an effective spin-$1/2$ or, as it is nowadays often referred
to, a qubit) in a dissipative environment, 
for this hallmark problem appears to describe a variety of
physically different (yet, formally related) situations \cite{Leggett}.

More recently, the attention has been drawn to this problem's further extension
where $N$ coupled two-level systems would be subject to a dissipative 
(possibly, non-uniform) environment. This interest was largely motivated 
by the relevance of the problem in question for practical
implementations of the ideas of quantum information processing.

However, despite the previous (mostly numerical and largely
limited to weak coupling) analyses of the problem 
of $N=2$ noisy qubits \cite{Shoen}, not much of a consideration has been given
to the assessment of a possibility of reducing the 
environmentally induced decoherence by means of properly choosing the parameters of
individual qubits and/or by virtue of permanent pairwise qubit interactions
which (whether desired or not) would be unavoidably present in any realistic setup.

As one step towards developing a systematic (as opposed to heuristic)
approach to addressing this kind of issues,
in the present work we explore a formal connection between the spin-boson model of interacting
qubits coupled to a generic (multi-component and/or (sub)Ohmic) bosonic bath
and some of the recently studied models of (non)magnetic impurities in strongly correlated
electron systems.

Such analysis would be incomplete without investigating the intermediate-to-strong coupling regime
where the customary Fermi's Golden rule-based estimates 
fail, and a more elaborate approach is needed.
While, at the first sight, a strong-coupling behavior may not seem to be
immediately relevant to the previously proposed designs of a practical quantum register, 
it should be noted that experimentally measured 
decoherence rates routinely exceed their best estimates obtained in the framework of the weak-coupling 
Bloch-Redfield and related approximations. 
Moreover, even in the case of a single qubit, the recent non-perturbative analyses
of the strong coupling behavior as well as the effects
of non-Markovian and/or structured environments have already resulted 
in a number of potentially interesting implications for quantum computing \cite{Costi}. 

Regardless of the qubits' physical make-up, the dynamics of $N$ such species is usually
described by the spin-boson Hamiltonian 
\be
H^{(N)}_{SB}=\sum_{a,i}{S}^a_i({B}^a+\gamma^ah^a_i)
+\sum_{a,ij}I_{ij}^a{S}^a_i{S}^a_j+\sum_k\omega_kb^\dagger_kb_k
\ee
The random field 
$h^a_i=\sum_ke^{i{\vec k}{\vec r}_i}{\lambda^a_k}(b^\dagger_k+b_{-k})/{\sqrt {\omega_k}}$
represents a generic multi-component bosonic bath composed of
$D$-dimensional propagating modes with the dispersion $\omega_k=vk$
and described by the correlation function (hereafter $R=|i-j|$
is a distance between the $i^{th}$ and
$j^{th}$ qubits and $\Lambda$ is an upper cutoff of order the bath's bandwidth)
\be
<h^a_i(t)h^b_j(0)>\propto{\delta^{ab}\Lambda^\epsilon\over [t^2-(R/v)^2]^{1-\epsilon/2}}
\ee
The variable parameter $0<\epsilon<1$ controls the bath's spectral density 
$\rho(\omega)=\sum_k(|\lambda^a_k|^2/{\omega_k})\delta(\omega-\omega_k)
\propto\Lambda^\epsilon\omega^{1-\epsilon}$, thus
allowing one to study the entire range of different (sub)Ohmic environments
within the same unifying framework.

Recently, a new reincarnation of the Hamiltonian (1)
has emerged under the name of the Bose Kondo model
in the theory of (non)magnetic impurities in strongly correlated systems, such  
as heavy fermions, Mott insulators, and high-$T_c$ cuprates \cite{Demler}.

In the case of a single qubit ($N=1$), the Hamilton (1)
gives rise to the renormalization group (RG) equations which,  
to the first order in the dimensional regularization parameter $\epsilon=3-D\ll 1$, read
$$
{dg_\parallel\over dl}=(\epsilon -\sum_{a=1,2}g_{\perp,a})g_\parallel
$$
$$
{dg_{\perp,a}\over dl}=(\epsilon-
\sum_{b\neq a}g_{\perp,b}-g_\parallel-{\vec B}^2)g_{\perp,a}
$$
\be
{d{B}_a\over dl}=(1-\sum_{a=1,2}g_{\perp,a}){B}_a
\ee
In these equations which generalize the analysis of Ref.\cite{Demler} to the case 
of a finite uniform field $\vec B\neq 0$, $l=\ln\Lambda/\omega$ is the energy-dependent renormalization scale.
The longitudinal $g_\parallel=({\vec \gamma}{\vec B})^2/B^2$
and the transverse $g_{\perp,a}$
($\sum_{a=1,2}g_{\perp,a}={\vec \gamma}^2-({\vec \gamma}{\vec B})^2/B^2$)
qubit-bath couplings are defined with respect to the direction of ${\vec B}=(0,0,{\tilde \Delta})$.

Eqs.(3) feature a variety of fixed points that can be classified according to the number of
effectively equal components of the vector $\vec g$. Namely, we find 
unstable $SU(2)$- ($g_{\perp,a}=g_\parallel$) and $XY$- 
($g_{\perp,a}=g$, $g_\parallel=0$) invariant 
fixed points, while the only stable one is a pair of
Ising fixed points ($g_\parallel=g_{\perp,1}=0$ or $g_\parallel=g_{\perp,2}=0$).
Thus, in a striking contrast with the conventional (Fermi) Kondo model \cite{Leggett},
an anisotropy of the bosonic couplings does not renormalize away, but instead tends to increase.

In the Ising case, the coupling to a one-component
Ohmic bath gives rise to the conventional Kosterlitz-Thouless (KT) transition that occurs
at $g_{I}=1$, ${\tilde \Delta}_{I}=0$ and separates the regime where a sufficiently
small initial ${\tilde \Delta}$ continues to decrease ($g>1$)
from that where it instead grows ($g<1$). In the language of the Fermi Kondo model
(see below), these two regimes correspond to the ferromagnetic (FM) Kondo effect 
and the usual antiferromagnetic (AFM) Kondo screening, respectively.

The behavior of the running RG variable
${\tilde \Delta}(l)$ should not, of course, be confused with that
of the renormalized splitting between the qubit's energy levels $\Delta(l)={\tilde \Delta}(l)e^{-l}$ 
which still decreases (albeit, in different ways) in both cases.
In the extensively studied AFM regime of the Ohmic problem, the latter
is given by the well-known solution 
\be
\Delta_*\sim\Lambda(\Delta_0/\Lambda)^{1/1-g}
\ee 
of the self-consistent equation ${\Delta}_*=\Delta(l=\ln\Lambda/\Delta_*)$ \cite{Leggett}.
 
In the case of a sub-Ohmic ($\epsilon>0$) bath and in the vicinity of the Ising fixed point, the RG flow
decribed by Eqs.(3) is characterized by a constantly increasing effective coupling $g(l)$ and 
a vanishing (possibly, after some temporary increase) ${\tilde \Delta}(l)$, 
provided that its bare value ${\Delta}_0$ falls below the separatrix which connects the unstable fixed
point at $g_{I}=1$, ${\tilde \Delta}_{I}=\epsilon^{1/2}$ to the trivial one at 
the origin of the $g-{\tilde \Delta}$ plane. 

Such a behavior signals a total loss of coherence and a complete qubit's
localization in one of the two degenerate states (overdamped regime), in agreement
with the customary expectations \cite{Leggett}. 
However, for ${\Delta}_0\gtrsim\Lambda [2ge^{1/2-g}]^{1/\epsilon}$ 
the solution of Eqs.(3) shows that coherent damped oscillations with a frequency
${\Delta}_*\sim\Lambda (2g)^{1/\epsilon}$ might still be possible, thus supporting the earlier prediction
made in Ref.\cite{Mielke}.

In contrast to the Ising case, the coupling to a multi-component ("non-abelian")
Ohmic bath may give rise to a markedly different behavior.
In both cases of the $XY$- and $SU(2)$-symmetrical couplings, our solution of Eqs.(3)
shows that the standard KT fixed point is absent, and the renormalized level splitting continues to grow 
(although it may temporarily descrease at $g(l)\sim 1$), thus 
enabling the qubit to avoid localization.

Furthermore, in the "weakly sub-Ohmic" ($0<\epsilon\ll 1$) case a new unstable 
fixed point emerges at $g_{XY}=\epsilon$ ($g_{SU(2)}=\epsilon/2$) for the $XY$- ($SU(2)$-)
symmetrical couplings and ${\tilde \Delta}_{XY,SU(2)}=0$. 
Notably, this non-trivial fixed point governs the regime of small $g(l)$ and ${\tilde \Delta}(l)$
which is of a primary importance for quantum computing-related applications of Eq.(1).

In the latter case, contrary to the naive expectation drawn from the weak-coupling analysis,
the coupling $g(l)$ may first increase, before succumbing to the continuing 
growth of ${\tilde \Delta}(l)$ and reverting to the opposite, decreasing, behavior,
provided that the initial values of the RG variables satisfy the condition
$max[g_0,{\tilde \Delta}_0^2]\lesssim\epsilon$.

Besides, Eqs.(3) indicate that for a strongly sub-Ohmic bath ($\epsilon>\epsilon_*=1/2$) 
yet another unstable $XY$-symmetrical fixed point can emerge at $g_{XY,II}=1/2$ and 
${\tilde \Delta}_{XY,II}=(\epsilon-\epsilon_*)^{1/2}$,
while the fixed point at $g_{XY,I}=\epsilon$ and ${\tilde \Delta}_{XY,I}=0$ 
would then become stable.
Considering the fact that the corresponding threshold value $\epsilon_*$ is quite 
large, this prediction may turn out to be spurious. 
Nevertheless, from a general viewpoint, it is
conceivable that even a multi-component (yet, sufficiently strongly
sub-Ohmic) bath could still be capable of causing a complete localization of the qubit
for small ${\tilde \Delta}_0$.

The above predictions for the RG flow of the couplings and the effective level splitting 
have a direct bearing on the properties of such observables as the qubits' correlation functions
$<{\{}S^a_i(t),S^a_j(0){\}}>$ and the corresponding dynamical spectral functions
\be
\chi_{ij}(\omega)=Im\sum_{n,m}(\rho_{nn}-\rho_{mm})
{<n|S^a_i|m><m|S^a_j|n>\over \omega-E_n+E_m+i\Gamma_{nm}}
\ee
given by the sums over renormalized and broadened 
energy levels $|n>$ of the interacting noiseless qubits
which are weighted with the diagonal elements of the
equilibrium density matrix $\rho_{nn}=|n><n|$.
 
The analysis of Eq.(5) reveals that
for non-interacting qubits the coherent (single-qubit) behavior manifests itself
through the presence of inelastic peaks at $\omega=\pm\Delta_*/2$
in the normalized spectral density $\chi_{ij}(\omega)/\omega^{1-\epsilon}\propto\delta_{ij}$
\cite{Costi,Mielke}. The width of these peaks is controlled by the decoherence rate
\be
\Gamma\propto g_*\rho(\Delta_*)\coth(\Delta_*/2T)
\ee
which increases as $\Gamma\sim g_*T(\Lambda/\Delta_*)^\epsilon$ with increasing
$g_*$ and decreasing $\Delta_*\lesssim T$.
As the qubits lose their coherence, the spectral weight gets transferred from the inelastic
peaks to the small energies ($\omega\approx 0$),
which behavior constitutes the decohering effect of the environment. 
The onset of complete localization is usually preceeded by exponential relaxation
which separates the former regime from that of the partly coherent damped oscillations
\cite{Leggett}. 

It is worth reminding that different probes may signal the loss of coherence at
different critical couplings. In the extensively studied Ohmic case, such a difference 
is examplified by the juxtaposition of the 
single-qubit average $<S^a_i(t)>$ and the auto-correlation function 
$\chi_{ii}(w)\coth\omega/2T$ whose coherent peaks get 
washed out at, respectively, $g=1/2$ (Toulouse point) \cite{Leggett} 
and $g=1/3$ \cite{Saleur}, both critical values being lower than the beforementioned 
estimate $g_I=1$.

In light of the above mentioned possibility of a non-monotonic behavior of $g(l)$ 
near the $SU(2)$- and $XY$-symmetrical fixed points at
small $g_0$ and ${\tilde \Delta}_0$, the initial increase of the coupling strength
may give rise to the situation where the inelastic peaks in the spectral function (5)
might temporarily become rather broad before they can narrow down at still lower energies.

The above findings strongly suggest that
by operating a noisy qubit in the vicinity of one of the $XY$- and $SU(2)$-symmetrical fixed points
one can better retain its quantum coherence.

One could argue, however, that for non-interacting qubits
the intrinsic instability of the model (1) towards developing the Ising-like anisotropy
may hinder any possibility of taking advantage of the 
greater robustness of quantum coherence found at the high-symmetry fixed points.
To this end, below we show that the loss of coherence caused
by the outward flow of the single-qubit RG trajectory away from a desired operating point
can be thwarted by inter-qubit couplings
which provide for an extra protection against decoherence.

In fact, some exchange-like interaction between the qubits is necessarily generated by the qubit-bath
couplings themselves. The instantaneous part of this (generally, retarded and FM-like)
interaction which arises in the course of integrating 
over the bosonic modes down to the energy scale $\omega\sim v/R$
\be
I^a_{B}=-\sum_k{\lambda^a_k\lambda^b_{-k}\over \omega^2_k}
e^{i{\vec k}({\vec r}_i-{\vec r}_j)}\propto -\delta^{ab}{g\Lambda^\epsilon/R^{1-\epsilon}}
\ee 
combines together with the direct inter-qubit coupling into the effective parameters $I^a_{ij}$
introduced in Eq(1).

It can be easily seen that a sufficiently strong FM exchange ($I\to -\infty$) forces a pair of 
qubits into a triplet state which then follows an effective $S=1$ Bose Kondo scenario.
In the opposite, strongly AFM, limit ($I\to +\infty$)
the two qubits get locked into a singlet state,
which prevents them from any unwanted entanglement with the environment.
According to Eq.(7), however, the latter regime can 
not be attained in the absence of a sufficiently strongly AFM direct coupling between the qubits.

At intermediate values, $I$ provides a cutoff for the RG flow
which now terminates at $l_I=\ln\Lambda/|I|$ before reaching the 
strong coupling limit, thereby resulting in a larger effective 
$\Delta_*=\Delta(l_I)$ that determines the position and the width of the coherent peaks.

In the case of the 3D Ohmic bath, a further insight can be gained
from the previously established correspondence between the $N=2$ Hamitonian (1) and 
the anisotropic two-impurity Fermi Kondo (TIFK) model 
$$
H^{(2)}_{TIFK}=-iv\sum_{i=1,2}\int^\infty_{-\infty}dxc^\dagger_{i}\partial_xc_{i}+\sum_aI^a{S}^a_1{S}^a_2
$$
\be
+\sum_a[J^a_{+}S^a_+(c^\dagger_{1}\sigma^ac_{1}+c^\dagger_{2}\sigma^ac_{2})+
\ee
$$
J^a_{-}S^a_-(c^\dagger_{1}\sigma^ac_{1}-c^\dagger_{2}\sigma^ac_{2})
+J^a_{m}S^a_+(c^\dagger_{1}\sigma^ac_{2}+c^\dagger_{2}\sigma^ac_{1})]
$$
where the auto-correlation function of the 1D spin-$1/2$ fermions with the Fermi momentum $k_F$ is
$<c_{i}^\dagger(t) c_{j}(0)>\propto\delta_{ij}/t$, while the Kondo couplings 
\be
J^a_+=J^a, ~~~J^a_m=J^a{\sin(k_FR)\over (k_FR)},~~~ J^a_-=J^a{\sqrt {1-{\sin^2(k_FR)\over (k_FR)^2}}}
\ee
between $S^a_{\pm}=S^a_1\pm S^a_2$ and the even, odd, and mixed bilinear 
combinations of the fermion fields $c_{1,2}=(c_e\pm c_o)/{\sqrt 2}$ 
at the two qubits' locations are given in terms of
the exchange constants of the single-impurity anisotropic Fermi Kondo model \cite{Gan}
(which, in turn, are related to the parameters of the $N=1$, $\epsilon=0$
Hamiltonian (1): $g=(1-(2/\pi)\tan^{-1}(\pi J^{\parallel}/4\Lambda))^2$ and
${\tilde \Delta}=J^{\perp}$ \cite{Leggett}).

In contrast to its bosonic counterpart (7), the RKKY-like inter-qubit coupling
mediated by the fermionic bath behaves
as $I_{F}\propto(J^2/\Lambda)((2k_FR)\cos(2k_FR)-\sin(2k_FR))/(2k_FR)^4$. 
Therefore, unless one chooses to neglect any environmentally
induced contributions to the exchange coupling $I$ altogether,
the strict correspondence between TIKM and the $N=1$ model (1)
can only hold for small inter-qubit separations, $R\ll 1/k_F$ (see below).

In the presence of the particle-hole symmetry,
the two-channel TIKM (8) possesses an unstable fixed point at 
a critical (AFM) value $I_*$ which is set by the single-qubit Kondo scale $T^{(S=1/2)}_K$.
This critical point, which in the $SU(2)$-symmetrical 
case ($T^{(S=1/2)}_K=\Lambda\exp(-\Lambda/J)$) occurs at $I_*\approx 2.5T^{(S=1/2)}_K$, 
gets replaced by a crossover, if the particle-hole symmetry is broken \cite{Gan}.

In the FM regime ($I\lesssim -T^{(S=1/2)}_K$), the Kondo screening of
the composite spin $S=1$ is characterized by
the scale $T_K(I)\sim T^{(S=1/2)}_K(T^{(S=1/2)}_K/|I|)^{\eta}$ which interpolates between 
$T^{(S=1/2)}_K$ and $T^{S=1}_K=T_K(I\lesssim -\Lambda)$. 
The non-universal exponent $\eta$ depends on the exchange 
anisotropy and assumes the value $\eta=2$
in the $SU(2)$-invariant case ($T^{(S=1)}_K=\Lambda\exp(-2\Lambda/J)$).

The above behavior pertains to the two-channel model which describes a pair of
well separated qubits ($R\gg 1/k_F$), whereas at small separations the term in (8)
proportional to $J^a_{-}\ll J^a_{+}\approx J^a_{m}$ can be 
disregarded, and Eq.(8) reduces to the
single-channel TIFK model formulated solely in terms of the even combination
$c_e$ of the fermion orbitals which is coupled to the total spin 
${\vec S}_+$. Remarkably, the two- and single-channel TIFK models appear to
describe the two opposite limits which correspond to 
independent and collective decoherence, respectively.

For a pair of identical qubits ($J_1=J_2$), the single-channel TIKM undergoes 
a simple first order transition (level crossing), as $I$ is 
tuned past its critical value $I_*$,
from the underscreened $S=1$ model at $I<I_*$ to the two-qubit singlet state 
which is completely decoupled from the bath at $I>I_*$ \cite{DQD}. 

If, however, the qubits are different ($J_1\neq J_2$), 
then, as $I$ increases past $I_*$, the more strongly coupled qubit
first gets screened at $T^{(I)}_K=\Lambda\exp(-\Lambda/max[J_1,J_2])$, followed by 
the KT-type transition (which now occurs regardless of the presence of the
particle-hole symmetry) at $T^{(II)}_K=max[I-I_*, T^{(I)}_K\exp(-T^{(I)}_K/I-I_*)]$.

The above conclusions, which were drawn in the case of the $SU(2)$-invariant
couplings, apply not only to qubits represented by physical (electron or nuclear) spins
but also to the effective qubit operators which represent, e.g., 
electron states in two-level single or lateral/vertical double quantum dots. 
Such operators have been routinely introduced in the Anderson or 
resonant level models where the genuinely $SU(2)$-invariant exchange 
couplings are generated from the electron tunneling amplitudes
by the Schrieffer-Wolff transformation.

A comparison between several different kinds of the
Ising-type qubit interactions studied in the context of the $N=2$ problem \cite{Shoen}
indicates that a pair of qubits can better retain their coherence
if the interaction term commutes with the qubit-bath coupling operator. 
In this regard, the Heisenberg exchange, 
which not only commutes with the overall qubit-bath coupling
in the case of collective decoherence but also provides for the strongest 
lifting of the singlet-triplet degeneracy, might offer the best available option for reducing
decoherence. 

We note, in passing, that spin-rotationally invariant couplings 
may not be easily realizable for some of the pseudo-spin qubits.
Among such examples are the Josephson junction qubits whose
Hamiltonians, while being potentially well-controllable,
generally would have no particular symmetries at all.
Therefore, it is conceivable that future designs of practical solid-state qubits
will need to optimize between the somewhat
contradictory requirements of efficient control and robust coherence.

The formal analogy between the problems of noisy qubits  
and quantum impurities established in this Letter can be pursued even further.
In particular, the ${\vec B}\neq 0$ counterpart of the previously studied Fermi-Bose Kondo
model \cite{Demler} can describe a system of qubits coupled to one Ohmic (represented by
a fermionic) and the other sub-Ohmic (bosonic) bath.
Such a situation occurs, e.g., in the Josephson qubits where the effects of both
the Nyquist ($\epsilon=0$) and $1/f$ ($\epsilon=1$, if treated as approximately Gaussian) 
noises which are caused, respectively, by fluctuating 
currents and background charges often need to be considered on equal footing. 

The fact that the only stable fixed points of the mixed Fermi-Bose model 
are the pure Fermi and pure Bose ones \cite{Demler} suggests a possibility
of effectively blocking off a more harmful type of coupling, while
tackling the remaining one with, e.g., error-correction techniques.

To summarize, in the present work we exploited a formal similarity between the problem 
of interacting qubits in (sub)Ohmic dissipative environments
and the recently studied models of quantum impurities in strongly correlated systems.
This insight enabled us to formulate a number of concrete recommendations for achieving the
optimal (coherence-wise) regime for operating a noisy multi-qubit quantum register. 

Firstly, we investigated the RG properties of the (sub)Ohmic single-qubit model (1)
in the vicinity of its fixed points of different symmetries. In the course of this
analysis, we discovered that for non-interacting qubits 
and a "weakly-sub-Ohmic" bath ($0\leq\epsilon<\epsilon_*$), 
the best possible conditions for preserving coherence would require 
the most spin rotationally-symmetrical qubit-bath couplings $\gamma^a$.

Secondly, we ascertained the possible 
benefits of permanent interactions $I^a_{ij}$ between the elements of a multi-qubit array
for attaining the most coherence-friendly 
regime and protecting the qubits' quantum memory during the idling 
periods between consecutive gate operations.
Specifically, we found that it would be advantageous to tune the (preferably, $SU(2)$-symmetrical) 
pair-wise inter-qubit coupling $I$ close to (yet, smaller than) the critical value $I_*$
of the TIFK model, thereby causing an incipient singlet 
formation and concomitant quenching of the Kondo screening.

Lastly, the Kondo physics-conscious approach to 
the engineering of robust qubit Hamiltonians 
might prove instrumental in solid state implementations of quantum information processing                                                  
where, unlike in liquid-state NMR or trapped-ion designs,                                                      
such active noise suppression techniques as dynamical decoupling/recoupling                                    
schemes may not be readily available.    
Therefore, we believe that this work will further spur the ongoing cross-fertilization between the
developing theory of quantum information and such
well-established topics in materials theory as Kondo physics of heavy fermions and quantum dots.

The author acknowledges valuable discussions with S. Girvin, A. MacDonald and S. Das Sarma.
This research was supported by ARO under Contract DAAD19-02-1-0049
and NSF under Grant DMR-0071362.

{\it Note added}: In the case $n=1$ and $\epsilon>0$, Eq.(2) of this 
work (first made available as a preprint [cond-mat/0305510]) and the conclusions 
regarding the possibility of a delocalization transition for
a sufficiently large $\Delta$ were recently confirmed in Ref.\cite{Vojta}.

\wt
\end{document}